\documentclass[letterpaper,twocolumn,pre,superscriptaddress,showpacs]{revtex4}

\usepackage{amsmath}
\usepackage{amssymb}
\usepackage{graphicx}
\usepackage[usenames,dvipsnames]{xcolor}
\usepackage{booktabs}
\usepackage{microtype}
\usepackage{hyperref}

\newcommand{\AkIJ}{A^{(k)}_{IJ}}
\newcommand{\BkIJ}{B^{(k)}_{IJ}}
\newcommand{\WkIJ}{W^{(k)}_{IJ}}

\hypersetup{colorlinks=true,citecolor=Red,linkcolor=Blue,urlcolor=Black}
\AtBeginDocument{
	\heavyrulewidth=.08em
	\lightrulewidth=.05em
	\cmidrulewidth=.03em
	\belowrulesep=.65ex
	\belowbottomsep=0pt
	\aboverulesep=.4ex
	\abovetopsep=0pt
	\cmidrulesep=\doublerulesep
	\cmidrulekern=.5em
	\defaultaddspace=.5em
}

\begin{document}

%
%
%
\title{Targeting Multiple States in the Density Matrix Renormalization Group with The Singular Value Decomposition}

\author{E.~F.~D'Azevedo}
\affiliation{Computer Science and Mathematics Division, Oak Ridge National Laboratory, %
	\mbox{Oak Ridge, Tennessee 37831}, USA}

\author{W.~R.~Elwasif}
\affiliation{Computer Science and Mathematics Division, Oak Ridge National Laboratory, %
	\mbox{Oak Ridge, Tennessee 37831}, USA}

\author{N.~D.~Patel}
\affiliation{Department of Physics, The Ohio State University, Columbus, \mbox{Ohio 43210}, USA}

\author{G.~Alvarez}
\affiliation{Computational Sciences and Engineering Division %
	 and Center for Nanophase Materials Sciences, Oak Ridge National Laboratory, %
 \mbox{Oak Ridge, Tennessee 37831}, USA}

\date{\today}
\pacs{71.10.Fd, 71.27.+a, 02.70.Hm, 79.60.-i}

\maketitle

\section{abstract}
In the Density Matrix Renormalization Group (DMRG), multiple states must be included in the density
matrix when properties beyond ground state are needed, including temperature dependence,
time evolution, and frequency-resolved response functions. How to include these states in the density
matrix has been shown in the past. But it is advantageous to replace the
density matrix by a singular value decomposition (SVD) instead, because of improved performance, and
because it enables multiple targeting in the matrix product state description of the DMRG. 
This paper shows how to target
multiple states using the SVD; it analyzes the implication of local symmetries, and discusses
typical performance improvements using the example of the Hubbard model's photoemission spectra
on a ladder geometry.

\section{Introduction}

The density matrix renormalization group (DMRG) \cite{re:white92, re:white93} finds the
ground state of local Hamiltonians, such as those that appear in condensed matter theory.
The computational complexity of DMRG grows as a polynomial in the number of lattice sites, if the local
Hamiltonian is formulated in a one dimensional chain.
Moreover, for quasi one dimensional systems, the computational complexity grows
as a polynomial in the length of the long dimension, such that
the DMRG can efficiently truncate all local Hamiltonians on these lattices. The error made
by the DMRG can be estimated, and it converges to the exact result as more and more states are kept.
To perform the truncation the lattice is divided in two disjoint parts: a system and an environment.

In conventional DMRG, the reduced density matrix $\rho^S$ of the system with respect to the environment
is computed using only the ground state vector, and then $\rho^S$ is diagonalized. The eigenvectors $W$
of the reduced density matrix are
used to effect a change of basis. In the new basis, the eigenvalues $d_\alpha$ of the reduced density
matrix determine the importance of each state, and states below certain importance are discarded.
If $A$ operates only on the system and $\bar{A}=W^\dagger A W$ in the new basis, then
\begin{align}
\langle A\rangle = \text{Tr}(A\rho^S) = \sum_\alpha \bar{A}_{\alpha,\alpha}d_\alpha &=
\sum_{\alpha<m}\bar{A}_{\alpha,\alpha}d_\alpha + E_A(m),\nonumber\\ \text{where} \,
E_A(m) = \sum_{\alpha>=m} \bar{A}_{\alpha,\alpha} & d_\alpha \le \bar{A}_{\text{max}}
\sum_{\alpha>=m}d_\alpha\nonumber
\end{align}
is the DMRG error.
Instead of changing basis using the eigenvectors of the reduced density matrix,
one may instead perform a singular value decomposition (SVD) of the ground state vector.
The square of the singular values then correspond to the eigenvalues of the reduced density matrix.
This SVD approach, though algebraically  equivalent, is faster, because the state
does not need to be ``squared'' to create a density matrix. 
Furthermore, the SVD approach blends well
in the matrix product state (MPS) formulation of the DMRG~\cite{re:rommer97, re:schollwock10}.

In applications of the DMRG to frequency dependent observables~\cite{re:hallberg95, re:kuhner99, re:jeckelmann02},
or to time-dependent
Hamiltonians \cite{re:white04, re:manmana05}, 
or to finite temperature properties \cite{re:feiguin05}, one needs multiple states instead of just a single state
like in ground state DMRG. For example, in the correction vector method for frequency dependent DMRG~\cite{re:kuhner99},
three states are needed to obtain the 
angle resolved photo-emission spectra and the Green functions: the ground state vector,
the vector resulting from the application of
the creation or destruction operator to the ground state, and the correction vector. To obtain the 
dynamical spin structure factor the vectors resulting from the application of spin operators to
the ground state are needed instead.
 
Yet the DMRG algorithm is not immediately applicable to arbitrary states, and was originally developed
to compute the ground state of the Hamiltonian instead.
This difficulty has been successfully overcome (see~\cite{re:schollwock05} and references therein)
by redefining the reduced density matrix
of the ``system'' or left block $\mathcal{L}$ as:
\begin{equation}
	\rho^S_{\alpha,\alpha'} =
	\sum_{\beta\in \mathcal{R}}\sum_f w_f \psi^{f\dagger}_{\alpha,\beta} \psi^f_{\alpha',\beta},
	\label{eq:rdensitymatrix}
\end{equation}
where $\alpha$ and $\alpha'$ label states in the left block $\mathcal{L}$,
$\beta$ those of the right block $\mathcal{R}$, and
$\{\psi^f\}_f$ is the set of states of the superblock $\mathcal{L}\cup\mathcal{R}$
that are needed for observations. The weights
$w_f$ are chosen to be non-zero.
The states $\psi^f$ are said to be \emph{targeted} by the DMRG algorithm. 
Because of their inclusion in the reduced density matrix, these states can be obtained with
a precision that scales similarly to that of the ground state in the static formulation of the DMRG.

But it is advantageous to replace the
density matrix construction by a singular value decomposition (SVD), as explained in the single state case.
In particular, it enables \emph{multiple state targeting} in the MPS formulation, and thus addition of MPSs
is not needed, neither is their consequent compression.
The SVD for multiple states is less trivial to perform:
The \emph{use of symmetries is far from obvious in the SVD approach with multiple states,} and requires
careful thinking of the equations. \emph{Parallelization over symmetry patches} speeds up the simulation further.

Subsection \ref{sec:svd} explains the singular value decomposition for multiple states, and
\ref{sec:patches} describes the acceleration brought about by symmetry patches.
Subsection \ref{sec:matrix_vector} overviews the related matrix-vector multiply, which is the
most time consuming sub-algorithm of the DMRG. Subsection \ref{sec:model} applies these algorithms
to a Hubbard model formulated on a two-leg ladder, and describes the spectral function obtained.
Subsection \ref{sec:performance} analyses the performance profile of the sub-algorithms,
emphasizing the improvements brought about by the replacement of the reduced density matrix calculation
by the SVD decomposition.
The conclusions provide an overview of implications for the MPS formulation
of the DMRG; pointers
to the free and open source computer programs used are also given.

\section{Theory}

\subsection{Singular Value Decomposition} \label{sec:svd}

We first briefly summarize the conventional and SVD approaches with one state, often the ground state.
We consider a lattice partitioned into a left part or ``system'' with $n_s$ states,
and a right part or ``environment'' with $n_e$ states. We consider the full lattice or \emph{superblock} states
labeled by $\alpha$ on the left, and $\beta$ on the right. 
If there is only one vector $\psi$ to target then the reduced density matrix of the left part is
$\rho_{\alpha', \alpha} = \sum_{\beta} \psi_{\alpha', \beta}\psi^*_{\alpha, \beta}$, which 
we shall represent as $\rho=\psi \psi^\dagger$. (The reduced density matrix of the right part is
similar, and will not be considered in what follows.)
One then fully diagonalizes $\rho=W^\dagger D W$, where $W$ are the eigenvectors of $\rho$ and
$D$ is the diagonal matrix of its eigenvalues.
It is known that we can replace the diagonalization of $\rho$ by the singular value decomposition (SVD)
of $\psi= USV^\dagger$, where it can be proved by substitution that $U$ coincides with the eigenvectors of $\rho$,
and that the square of $S$ contains in its diagonal the eigenvalues of $\rho$ \cite{re:schollwock10}. 

We now extend the single state approach to the 
targeting of multiple vectors $\psi^f$, $0\le f < F$.
The definition of the reduced density matrix is then $\rho=\sum_f w_f \psi^f \psi^{f\dagger}$, where the weights
$w_f > 0$, and the sum over $f$ runs over all the $F$ vectors that need to be targeted.
We define the vector $X_{f, \alpha; \beta}=\sqrt{w_f}\psi^f_{\alpha, \beta}$ such that $X$ can be thought of as a matrix
of $F\times n_s$ rows and $n_e$ columns; $\rho=XX^\dagger$ and
the replacement for the full diagonalization of $\rho$ is
then the SVD of $X$:
\begin{equation}
X_{f, \alpha; \beta} = \sum_{k, k'} U_{f, \alpha; k} S_{k, k'} V^\dagger_{k', \beta}.
\end{equation}
One can then see by substitution into the equality $\rho=XX^\dagger$ that $U$ contains the eigenvectors of $\rho$ and $S^2$ its eigenvalues.
We must now carefully develop these formulas to take into account \emph{symmetry patches} by
classifying the states by their local symmetry properties. This is needed to preserve symmetries and helpful
to speed up the computational simulation. Patching formulas are divided in two: (i)
organizing $\psi^f$ in patches to be ready for SVD,
and (ii) performing the actual SVD in patches.

\subsection{Symmetry Patches}\label{sec:patches}

The DMRG algorithm represents the full Hamiltonian $H$ as sum of 
Kronecker product of operators
\begin{equation}
H = H_L \otimes I + I \otimes H_R 
+ \sum_k C_L^k \otimes C_R^k,
\label{eq:hfull}
\end{equation}
where $C_L$ are matrices on the left block; these matrices vary from model to model:
for the Hubbard model they are the electron creation matrices, for the Heisenberg
model, these are the z-component of the spin $S_z$, and the matrices $S^+$ that increase
the z-component of the spin, and similarly for other models. Likewise, the counterpart
matrices on the right block are denoted by $C_R$.

\begin{figure}
	\centering
	\includegraphics[width=0.45\textwidth]{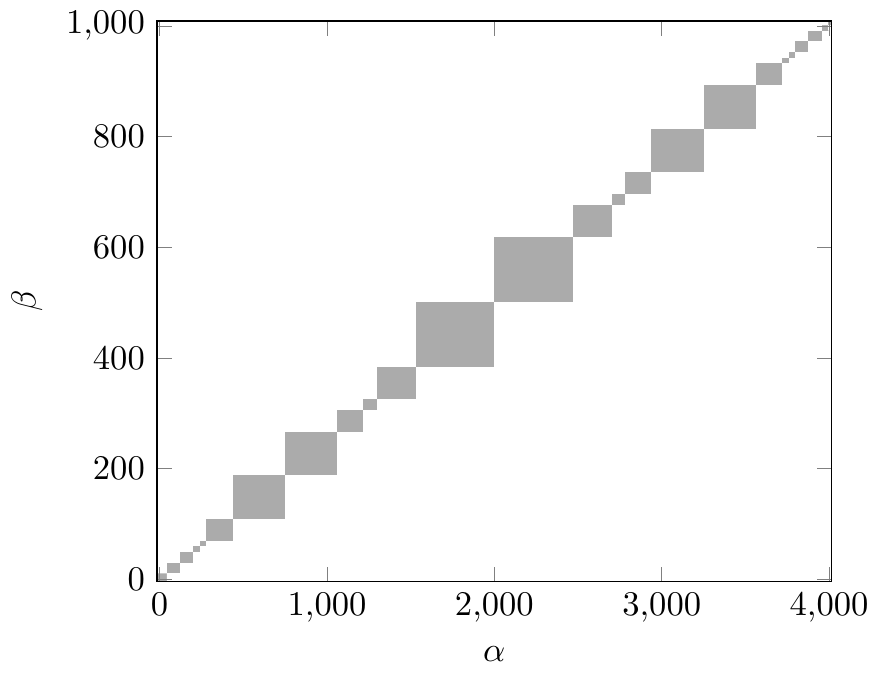}
	\caption{Example of a full vector $\psi_{\alpha, \beta}$ in the superblock,
		when reshaped as $1000 \times 4000$ matrix to
		highlight non-overlapping rectangular patches.
		The basis has been reordered such that $\psi_{\alpha, \beta}$
		becomes block diagonal, which is 
		always possible but not necessary.}
	\label{fig:patches}
\end{figure}

The states in each vector $\psi$ can be reorganized by grouping
them according to the left or right quantum number. 
For example,
if there are $n_s = 1000$ states on left or system and $n_e=4000$ states on right or environment, then
the $\psi$ vector can be reshaped as a $1000 \times 4000$ matrix.
The non-zero entries in $\psi$ can be grouped as rectangular ``patches'', such 
that  $\psi_{\alpha, \beta} = 0$ unless $q_\alpha + q_\beta = q_{\textrm{target}}$,
where $q_\alpha = (n_e(\alpha), S_z(\alpha))$
with $n_e(\alpha)$ the number of electrons of state $\alpha$, and $S_z(\alpha)$ the spin component $z$
of state $\alpha$.
This symmetry constraint would then lead to a patched matrix $\psi_{\alpha, \beta}$ that is
not necessarily block diagonal.
Figure~\ref{fig:patches} shows the case where the basis has been reordered such that
$\psi_{\alpha, \beta}$ becomes block diagonal, which is 
always possible but not necessary. 
Whether block diagonal or not, there is only one patch for each block row or block column, and
the
SVD of $\psi_{\alpha, \beta}$ can then be computed independently as the
SVD of each individual rectangular patch.

\subsection{SVD in Patches}


The DMRG algorithm described previously reshapes the lowest eigenvector $\psi$ produced
by the Lanczos algorithm  into a $L \times R$ rectangular matrix  to form the density matrix
$\rho = \psi^\dagger \psi$ where $L$ is the number of basis vectors in ``left'' and $R$
the number of basis vectors in ``right''. 
The eigen decomposition of $\rho$ is then used in a truncation
process to select the largest dominant eigen-pairs to be used in the
next step. 

The eigen decomposition can be more efficiently obtained by performing
singular value decomposition (SVD) of rectangular matrices over the patches in $\psi$.
Let us first observe that the matrix $\rho$ is block diagonal. The number of diagonal blocks
is the number of symmetry patches in the eigen-vector $\psi$.
If $\psi_i$ is the rectangular matrix for the $i^{\mbox{\scriptsize th}}$ 
patch, then
the $i^{\mbox{\scriptsize th}}$ diagonal block of $\rho$ is $\psi_i^\dagger \psi_i$.
(Do not confuse $\psi_i$ a symmetry patch of the single vector $\psi$, with $\psi^f$ 
	the $f$-th vector in the multiple target case.)
Note that the economy version of SVD of a $m \times n$ matrix $A$  gives
$ A = U *S *V^\dagger $ where 
$k \le \min(m,n)$ is the rank of matrix $A$;
matrix $U$ is $m \times k$ and has
 orthogonal columns and $U^\dagger*U = I_k$;
matrix $S$ is a $k \times k$ diagonal matrix that contains the singular values;
matrix $V$ is $n \times k$ and
has orthogonal columns $V^\dagger*V = I_k$ with 
 $I_k$ the identity matrix of size $k \times k$.
The right singular vectors are then the eigen vectors of $A^\dagger A$ and 
the singular values  are the positive square roots of the eigen-values,
\begin{equation}
A^\dagger A = (U S V^\dagger)^\dagger  (U S V^\dagger) 
      = V (S^\dagger S)  V^\dagger
~.
\end{equation}
Thus the truncation process can proceed by
independently computing the SVD of rectangular matrices over each
patch of $\psi$ concurrently.  The right singular vectors are the 
eigen-vectors; the eigen-values are the squares of singular values.
\emph{The eigen-pairs of the density matrix can thus be obtained but without explicitly forming $\rho$.}

\subsection{Multiple Targets}\label{sec:multiple_targets}

States included in the reduced density matrix of either system or environment are
said to be \emph{targeted.} In ground state calculations, the ground state
is often (but not always \cite{PhysRevB.72.180403})
the only state targeted. In order to apply DMRG to finite time or to finite
temperature or to finite frequency problems, other states need to be targeted.
We now give a brief summary of these applications. 
In each application or variant,
 states other than the ones mentioned below may be included in the reduced density
matrix or targeted.
For finite time, the time evolved states of a wave-packet need to be targeted.
For finite temperature, the infinite temperature state and the inverse-temperature-evolved 
state need to be targeted. For finite frequency, the correction vectors need to
be targeted. 

Without loss of generality, we consider $F=3$ states
included in the reduced density matrix
\begin{equation}
\rho = A_0^\dagger A_0 + A_1^\dagger A_1 + A_2^\dagger A_2,
\label{eqn:rhomult0}
\end{equation}
where we have absorbed the weights $w_f$ into the matrices such that $A_f\equiv \sqrt{w_f} B_f$.
Each matrix  $A_f$ is the same shape $m$ by $n,$ and $\rho$ is $n$ by $n.$
With $\rho$ so defined, the DMRG proceeds with its truncation procedure by using
the eigenvalues and eigenvectors of $\rho$.

The SVD version stacks the matrices into a bigger matrix $3m$ by $n,$
\begin{equation}
A_{\textrm{mat}}  = \left( \begin{smallmatrix}  A_0 \\ A_1 \\ A_2 \end{smallmatrix} \right)
\end{equation}
and uses the SVD decomposition of $A_{\textrm{mat}}$ to truncate the DMRG states.
It can then be verified that this process is equivalent to
the traditional process of using the eigenvalues and eigenvectors of $\rho$ given
by Eq.~(\ref{eqn:rhomult0}).
First, $A_{\textrm{mat}}^\dagger  A_{\textrm{mat}}$
is   $n \times 3m$ by  $3m \times n$  
and the result is  $n$ by $n$ and equal to
$A_0^\dagger A_0 + A_1^\dagger A_1 + A_2^\dagger A_2,$ which is $\rho.$
Next, the economy SVD version of $A_{\textrm{mat}},$ 
yields at most min$(3m, n)$ non-zero singular values.
If  $A_{\textrm{mat}} = U  S V^\dagger$,    then
\begin{equation}
        \rho = A_{\textrm{mat}}^\dagger A_{\textrm{mat}} 
             = V  (S^\dagger S)  V^\dagger
             ~,
\end{equation}
yields the information related to eigen decomposition of $\rho$;
$S$ is diagonal and so is
$S^\dagger S$. 

\subsection{Matrix-Vector Product}\label{sec:matrix_vector}

The matrix-vector multiply of $H$ with a vector in the superblock is the
most time-consuming sub-algorithm of the DMRG. Although not directly related to the
density matrix or the SVD of the targeted
states, we briefly review it here and in the supplemental \cite{re:supplemental} for completeness.

As mentioned in \ref{sec:patches}, the reorganization or grouping of states by quantum numbers to
identity the  rectangular patches
offers a significant computational advantage where
the target Hamiltonian can be further expressed as
sum of Kronecker products of small matrices.
The evaluation of the matrix-vector multiplication $H \psi$
can take advantage of
the properties of Kronecker products \cite{re:supplemental} and 
be efficiently evaluated  as many independent matrix-matrix multiplications.

Given a targeted quantum number, there are only a finite
number of admissible combinations of left and right quantum numbers that
forms the upper bound on the number of patches $N_p$.
This reorganization or grouping corresponds to a matrix block
partitioning of the left and right operator matrices. 
The evaluation of the matrix-vector multiplication $H \psi$
can be viewed as matrix operations 
$Y = C  X$ where $C$ is a $N_p$ by $N_p$ block
partitioned matrix.
Each submatrix $C[I,J]$ in the block partition can be expressed
as a sum of Kronecker products,
$C[I,J] = \sum_k  \AkIJ \otimes \BkIJ$.
Each $\AkIJ$ ( or $\BkIJ$) corresponds to
a left (or right) operator in Eq.~(\ref{eq:hfull}).
There is a corresponding matching
partition in vectors $X$ and $Y$ as
$Y[I] = \sum_J C[I,J]  X[J]$
for row partition index $I$.
The expression $C[I,J]  X[J] = \sum_k ( \AkIJ \otimes \BkIJ) X[J]$
can be evaluated as matrix-matrix multiplications~\cite{re:VanLoan00}
\begin{align}
C[I,J]  X[J] &= \left( \sum_k^K \AkIJ \otimes \BkIJ \right) X[J] \nonumber\\
&= \sum_k^K ( \BkIJ  X[J]  (\AkIJ)^t )  
= \sum_k^K ( \WkIJ  (\AkIJ)^t )\nonumber
\end{align}
where $\WkIJ  = \BkIJ  X[J]$, 
the segment $X[I]$ is reshaped
as a matrix of appropriate shape,
and $^t$ indicates matrix transpose.
Note that depending on the specific model or geometry and number of operators, 
some of the $\AkIJ$ or $\BkIJ$ matrices may be the identity matrix or
even be zero, i.e., there may be 
different number (perhaps even zero) of Kronecker matrices in $C[I,J]$ block.
Each evaluation of $Y[I]$ can proceed independently.
Moreover, in $C[I,J]$ there may  be 
multiple independent matrix-matrix multiplication
operations in computing $\WkIJ = \BkIJ  X[J]$ over index $k$.
Thus the reorganization or group into patches exposes
many independent matrix-matrix multiplication 
operations that can be evaluated  in parallel.

\section{Case Study: Spectral function of Hubbard ladders}

\subsection{The Model and Dynamic Observables} \label{sec:model}
To study the SVD of multiple states, we consider the paradigm of a
Mott insulator simulated with DMRG using the Hubbard Hamiltonian~\cite{re:hubbard63,re:hubbard64b}
\begin{equation}
	H=\sum_{i,j,\sigma} t_{ij} c^\dagger_{i\sigma} c_{j\sigma}
	+U\sum_i n_{i\uparrow} n_{i\downarrow}.
\end{equation}
The Hilbert space where $H$ acts is the tensor product of Hilbert spaces on each site of the lattice.
Each one-site Hilbert space has four states in its computational basis: empty, occupied with electron up,
occupied with electron down, and occupied with two electrons of different spin.
The operator $c^\dagger_{i\sigma}$ creates an electron on site $i$ with spin $\sigma$, and $n_{i\sigma}$
is a diagonal operator that multiplies the state by $1$ if there is an electron with spin $\sigma$ at
that site, or yields the zero of the Hilbert space if not.
The hopping matrix $t$ corresponds to a two-leg ladder, that is, $t_{i,j}=t_x=t_y$ if sites $i$ and $j$ are
neighbors on a two-leg ladder and $0$ otherwise. We take $t_x = t_y = 1$ as the unit of energy.
Ladders have been investigated as the next step beyond chains and as a proxy for the
fully two-dimensional lattice; ladders are also of interest in the simulation of high temperature superconductors.

To obtain the angle-resolved photoemission spectra $A(q, \omega)$ at momentum $q$ and frequency $\omega$,
the states $\psi^f$ in  Eq.~(\ref{eq:rdensitymatrix}) must
include \cite{re:kuhner99} the ground state (which ought to be computed first), $c_r|gs>$
 (the state resulting from the application of the 
destruction operator $c_r$ to the ground state), and the real and imaginary parts of the correction
vector $|cv(r, \omega, \eta)\rangle \equiv (z-H)^{-1} c_r |gs\rangle$, where $z=\omega + i\eta$, and $r$ is a given site.
(We do not use arrows to denote sites, even though sites belong to a ladder and are thus two dimensional.)
One then \emph{observes} $c^\dagger_{r'}$ so that $A^+(r, r', \omega) = \langle gs| c^\dagger_{r'} | cv\rangle$, and one also must
 add \cite{re:dargel11} the counterpart $A^-$ to obtain $A(r, r', \omega)$.

\begin{figure}
	\includegraphics[width=0.45\textwidth]{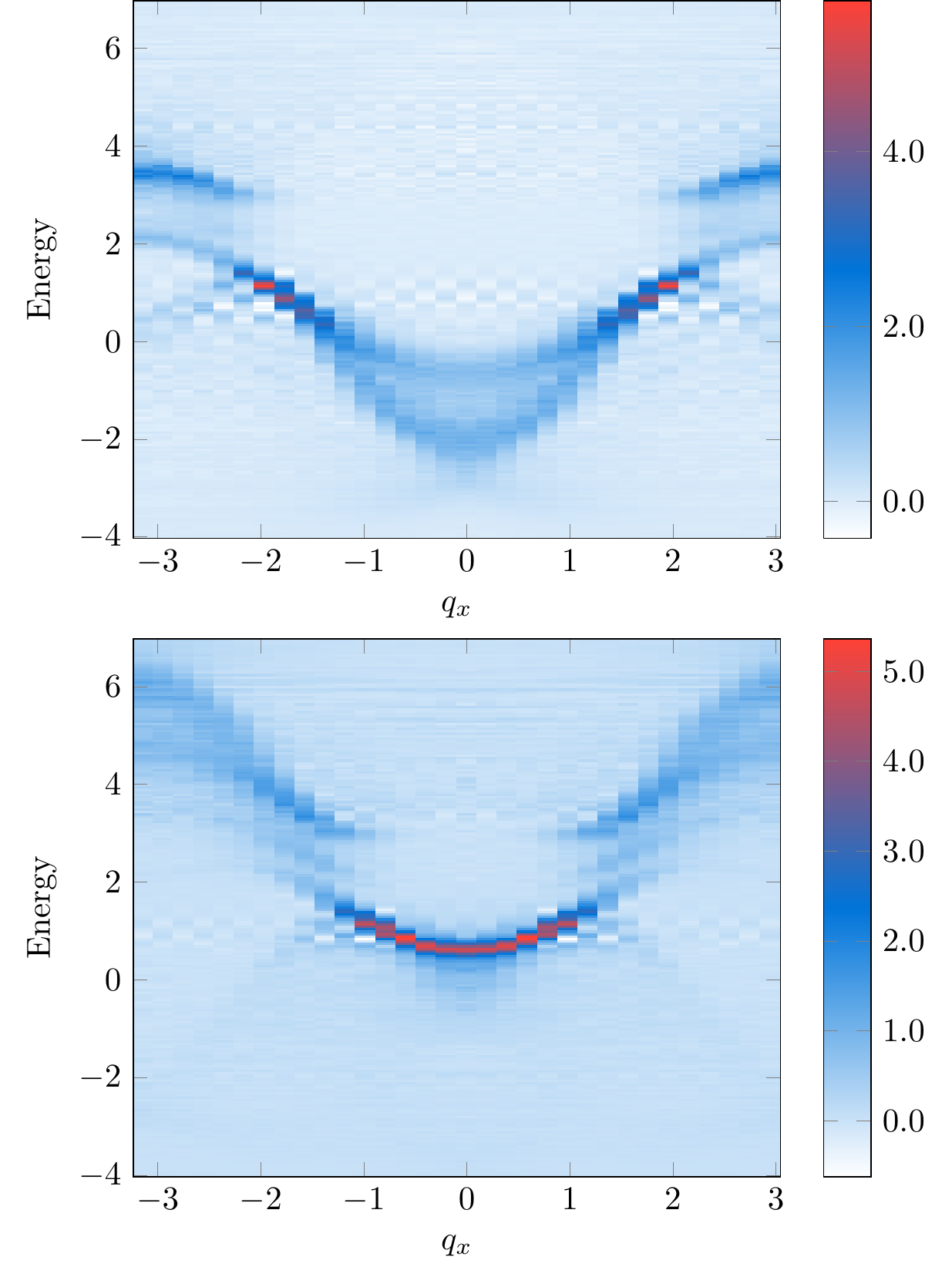}
	\caption{\label{fig:akw} (a) $A(q_x, q_y=0; \omega)$ of a 32$\times$2
		Hubbard ladder with $t_x=t_y=1$ as unit of energy, 
		$U=4$, and electronic $n=0.9375$ yielding
		a chemical potential equal to $1.20\pm0.05$. (b) Same as (a) for $q_y=\pi$}
\end{figure}

If the lattice were periodic and $A(r, r', \omega)$ depended only on the Euclidean distance between sites $r$ and $r'$,
then we could fix (say) $r$ and vary $r'$.
DMRG simulations are usually done with open boundary conditions in the long dimension, 
and ours is no exception. We use nevertheless an approximation
that we shall name ``center site approximation'' \cite{re:Noack1995}, where we
take one central site $r=c$ fixed, and vary the other site $r'$. This approximation accelerates our simulation because in one single
DMRG run we can obtain values $A(c, r', \omega)$ for all $r'$ at a fixed $\omega$. In parallel, we carry out the
needed simulations to obtain all frequencies $\omega$ of interest. We then perform a Fourier transform from space to
momentum $q$ to finally compute $A(q, \omega)$ of physical and experimental significance. 
Figure \ref{fig:akw} shows the result for both $q_y=0$ and $q_y=\pi$ on a $32\times2$ ladder
at $U=4$, with 4 holes for an electronic density of $n=0.9375$. Due to the ``center site approximation''
and the open boundary condition in the long direction of the ladder, the
imaginary part of the spectral function is slightly negative at frequencies where we would expect it to be zero,
a problem that we have not tried
to hide but show in the figure for explication purposes. 
If needed, such problems can be controlled either by finite-size scaling
or by using frequency filters.

Figure \ref{fig:akw} reproduces results presented in \cite{Riera1999}, now in a larger lattice at high accuracy.
The weights for the states included in the SVD is $0.3$ for each correction vector, $0.3$ for the $c_r |gs\rangle$
vector, and $0.1$ for the ground state vectors. (Other weight values could have been used; see the discussion in
\cite{re:kuhner99}.)
The chemical potential is at $\mu=1.20\pm0.05$.
The $q_y=0$ \emph{bonding band} shows high intensity 
at an energy near the chemical potential and at a momentum near $q_x = \pm 2$. 
A double
feature appears around momentum $q_x=0$, and toward $q_x=\pm \pi$ the spectral intensity appears 
at the high energy end of the spectrum ($\omega=3.5$).
The $q_y=0$ \emph{anti-bonding band} has intensity at the chemical potential 
and around momentum $q_x=0$, but most
intensity is at higher energies; the Hubbard band appears as expected at an energy $U=4$ above.

\subsection{Performance Profile} \label{sec:performance}

\begin{figure}
	\includegraphics[width=0.45\textwidth]{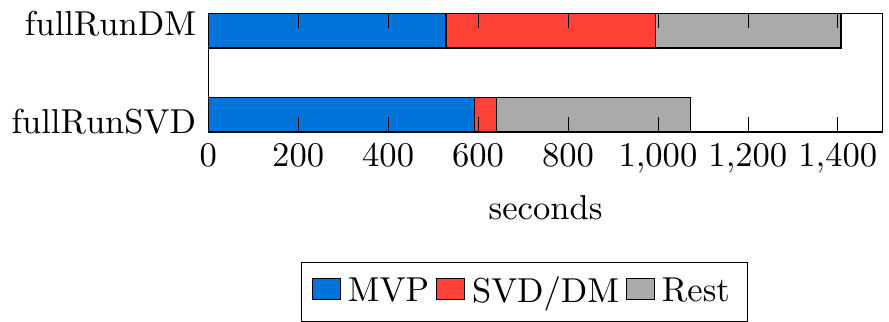}
	\caption{\label{fig:barchart} Wall times in seconds spent by the matrix-vector multiplication,
	the SVD or the DM depending on the type of sub-algorithm used, and the rest of the run, for
both the run done with SVD (fullRunSVD) and for the run done with the 
conventional density-matrix construction (fullRunDM).
The run for frequency $\omega=-2$ was chosen; other frequencies are similar.}
\end{figure}

The formation of the density matrix takes a part of the performance profile of a typical non-ground state DMRG run:
It accounts for approximately 40\% of the total run time.
(For ground state runs the density-matrix construction takes less percentage-wise \cite{NEMES20141570}.)
 When replaced by the SVD algorithm, 
this percentage reduces to only 5\% of
the total run time. Figure~\ref{fig:barchart} shows a bar-chart of the density matrix (DM) subalgorithm and its
replacement by SVD.
The figure indicates small variations in sub-algorithms not directly related to the SVD or DM, because
the states that are discarded vary even in identical runs due to the degeneracy of the eigenvalues of the density
matrix, at least to machine precision.
Nevertheless, figure~\ref{fig:barchart} illustrates the substantial gain in CPU performance that the use of 
SVD with symmetry patches brings about.
The rest of time is spent in other sub-algorithms, of which the construction
of Hamiltonian connections between ``system'' and ``environment'' at over 50\% of run time, is the most time consuming, as
we now explain.

In ground state DMRG, the most computational expensive kernel is the 
\emph{on-the-fly} computation of the Hamiltonian connections between system and environment
in order to perform Lanczos decomposition to obtain the ground state.
In the case of frequency dependent DMRG, the most computational expensive kernel is
the Lanczos tridiagonalization of the Hamiltonian, which again reduces to the
\emph{on-the-fly} computation of the Hamiltonian connections between system and environment.
As explained in reference \cite{nocera2016spectral}, the correction vectors \cite{re:kuhner99} 
are obtained in Krylov-space considering that
\begin{equation}
(z-H)^{-1} c |gs\rangle = (z-V^\dagger T V) ^{-1} c |gs\rangle,
\label{eq:krylov}
\end{equation}
where $z=\omega + i\eta$ with $\eta > 0$, $V$ the Lanczos vectors; there are $n_s$ Lanczos vectors,
each of size $n_b$. $T$ of $n_s$ rows and $n_s$ columns is the tridiagonal decomposition of $H$.
$H$ of $n_b$ rows and $n_b$ columns is too large to store in memory, and needs to be built on-the-fly.
We have found that $n_s=400$ Lanczos vectors are enough to study a $32\times2$ ladder.
We have kept at most $m=2000$ states for the DMRG, such that $n_b \le 4^2 m^2$ is the dimension of
the truncated Hilbert space; the factor $4$ arises because the one-site Hilbert space of the
Hubbard model is composed of 4 states.
\begin{figure}
	\centering{%
		\includegraphics[width=0.45\textwidth]{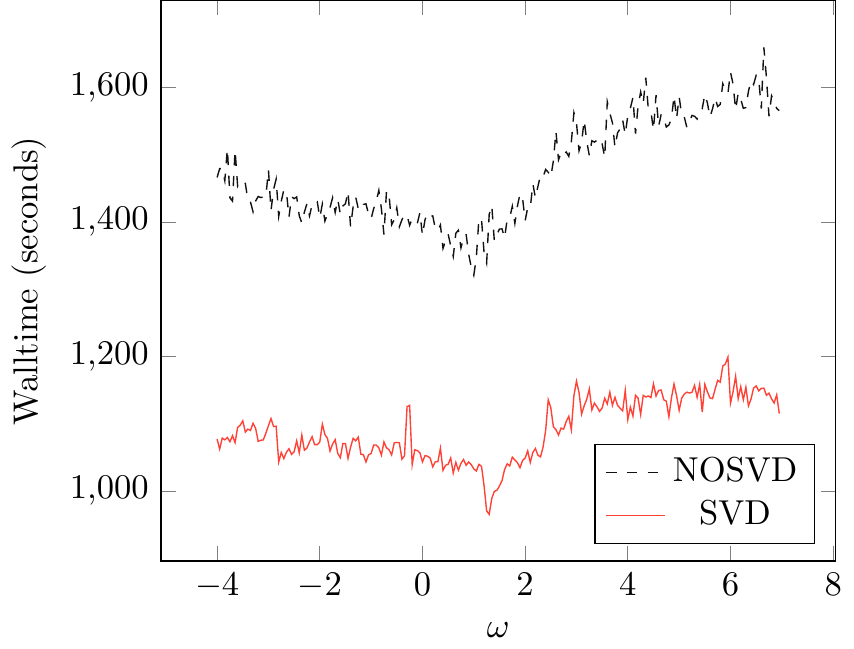}}
	\caption{\label{fig:walltimes} Walltimes in seconds for each frequency run $\omega$ using
		conventional DMRG (dashed line) and using SVD (solid line).}
\end{figure}

We have analyzed the performance of the model just described in the realistic
case of running the computer program in high performance computer systems,
and including the use of the GPU for the matrix-vector product algorithm.
Figure~\ref{fig:walltimes} shows
the total walltimes in seconds for a simulation using the traditional DMRG approach of building
the reduced density matrix, and using the SVD approach instead, for each frequency of interest.
For all frequencies, the SVD approach decreases total runtime by a factor of approximately 1.25; larger
frequencies take longer to converge in this Krylov-space algorithm as detailed in ref.~\cite{nocera2016spectral}.

Table~\ref{tbl:times} compares the same run when done on GPU and when done on CPU; only
the matrix-vector-product algorithm was run on GPU when indicated. The SVD algorithm is already
at 5\% of run time and it might not be beneficial to run it on the GPU due to memory overhead.
The results shown in the table should not be taken to infer that the GPU is linearly faster
than the CPU, because only the most computationally expensive sub-algorithm was
done on the GPU: the matrix-vector product. Other sub-algorithms, the SVD of the
vectors, the summation of sparse matrices, and the wave-function-transformation \cite{re:white96}
were done on the CPU. 
Data movement between the CPU and the GPU is another well-known limiting factor that must be
taken into account.

\begin{table}
\begin{tabular*}{\columnwidth}{@{\extracolsep{\stretch{1}}}*{5}{r}@{}} \toprule
	& \multicolumn{2}{c}{GPU} & \multicolumn{2}{c}{CPU} \\ \midrule 
		$\omega$ & fullRunSVD  & fullRunDM  & fullRunSVD  & fullRunDM  \\ \midrule 
	GS & 191 & 231 & 193 & 241 \\
	$-2$ & 1073& 1406 &1690 &1969\\
	$0$ & 1043 &1380& 1710& 1947\\
	$2$ &1048& 1402 &1757& 2014\\ \bottomrule
\end{tabular*}
\caption{\label{tbl:times}
	Wall times in seconds of typical runs depending on the use of the
SVD algorithm or the density matrix (DM) algorithm, for runs done using the GPU or the CPU
for the matrix-vector product algorithm. Note that the SVD or the density matrix diagonalization
was always done on CPU.}
\end{table}

\section{Conclusion}
Whether the MPS formulation or the conventional formulation of the DMRG is used, multiple
states must be targeted for observations beyond ground state.
The singular value decomposition helps both formulations. In the MPS formulation, targeting
multiple states replaces the addition of MPSs and their subsequent 
compression \cite{re:schollwock10} at the expense
of the maximum bond dimension $m$. In the conventional formulation the SVD 
replaces the more expensive density matrix sub-algorithm, substantially
reducing the time to solution.

Future work will apply the computational insights and theory explained in this paper
to the simulation of models on more than two leg ladders, of interest as a proxy to
the fully two-dimensional lattice. The real frequency spectral functions in these
models, of interest due to the existence of  angle-resolved
photo-emission spectroscopy and neutron scattering data, has mostly been inaccessible theoretically
and is only now been computed on large enough lattices and with enough precision to help
explain the transitions and interactions that cause the experimentally measured spectra.

All the theory and computation presented for real frequency applies straightforwardly
to finite temperature $T$ by replacing the initial state with the
infinite temperature state obtained from the ground state of entangler Hamiltonians
\cite{re:feiguin05, PhysRevB.93.045137} and the Lehman formulation in Eq.~(\ref{eq:krylov}) by
thermal evolution at ``imaginary time'' $\beta=1/T$. Likewise, real time evolution (when
using Krylov-space decomposition) will show
maximal computational cost in the matrix-vector product \cite{re:alvarez11}.
The computer programs used (including  \textsc{DMRG++} \cite{re:alvarez09}) are described
in the supplemental \cite{re:supplemental}, where details to enable reproduction of
the results presented here can also be found.

\section*{Acknowledgments}
The performance and algorithmic improvements have been supported by the 
scientific Discovery through Advanced Computing (SciDAC) program funded by
U.S. Department of Energy, Office of Science, Advanced Scientific Computing Research and Basic Energy Sciences,
Division of Materials Sciences and Engineering.
The case study was supported by the Laboratory Directed Research and Development
Program of ORNL.
Software development has been partially supported by the Center for Nanophase Materials Sciences,
which is a DOE Office of Science User Facility.
\bibliography{thesis}
\end{document}